%% file: short-v21.tex
\documentclass[english,prb,aps,amsmath,amssymb,amsfonts,floatfix,superscriptaddress,showpacs,twocolumn]{revtex4}
\usepackage[T1]{fontenc}
\usepackage[latin9]{inputenc}
\usepackage{verbatim}
\usepackage{float}
\usepackage{graphicx}

\makeatletter

\usepackage{epsfig}

\input{def.tex}

\def\beq{\begin{equation}}
\def\eeq{\end{equation}}

\def\LAs{$\rm{LaOFeAs}$~}
\def\LP{$\rm{LaOFeP}$~}

\makeatother

\usepackage{babel}

\begin{document}

\title{
Bandwidth and Fermi surface of Iron-Oxypnictides: covalency and sensitivity
to 
structural 
changes}

\author{Veronica Vildosola}

\affiliation{Centre de Physique Th\'{e}orique, \'{E}cole Polytechnique, CNRS, 91128 Palaiseau,
France}

\affiliation{Departamento de F\'{\i}sica, Comisi\'{o}n Nacional de Energ\'{\i}a At\'{o}mica (CNEA-CONICET),
Provincia de Buenos Aires, San Mart\'{\i}n, Argentina}

\affiliation{Japan Science and Technology Agency, CREST}

\author{Leonid Pourovskii}

\affiliation{Centre de Physique Th\'{e}orique, \'{E}cole Polytechnique, CNRS, 91128 Palaiseau,
France}

\author{Ryotaro Arita}

\affiliation{Department of Applied Physics, University of Tokyo, Tokyo
113-8656, Japan}

\affiliation{Japan Science and Technology Agency, CREST}

\author{Silke Biermann}

\affiliation{Centre de Physique Th\'{e}orique, \'{E}cole Polytechnique, CNRS, 91128 Palaiseau,
France}

\affiliation{Japan Science and Technology Agency, CREST}

\author{Antoine Georges}

\affiliation{Centre de Physique Th\'{e}orique, \'{E}cole Polytechnique, CNRS, 91128 Palaiseau,
France}

\affiliation{Japan Science and Technology Agency, CREST}

\begin{abstract}
Some important aspects of the electronic structure of the iron oxypnictides 
depend very sensitively on small changes in interatomic distances and bond angles within the iron-pnictogen
subunit.
Using first-principles full-potential electronic structure calculations, we investigate
this sensitive dependence, contrasting in particular \LAs and LaOFeP.
The width of the Fe-bands is significantly larger for LaOFeP, indicating a better metal and
weaker electronic correlations.  When calculated at their
experimental crystal structure these two materials have significantly
different low-energy band structure. The topology of the Fermi surface changes when going from \LP to LaOFeAs,
with a three-dimensional hole pocket present in the former case
transforming into a tube with two-dimensional dispersion.
We show that the low-energy band structure of \LAs evolves towards that of \LP as
the As atom is lowered closer to the
Fe plane with respect to its experimental position.
The physical origin of this sensitivity to the iron-pnictogen distance is the
covalency of the iron-pnictogen bond, leading to strong hybridization effects.
To illustrate this, we construct Wannier functions, which are found to have a large
spatial extension when the energy window is restricted to the bands with dominant iron character.
Finally, we show that the Fe bandwidth slightly increases as one moves along the rare-earth series in ReOFeAs and discuss the physical origin of this effect.
\end{abstract}
\maketitle

\section*{Introduction}

The recent discovery of superconductivity in electron-doped rare-earth oxyarsenides\cite{Kamihara-LOFA}
has generated a great deal of interest in the electronic structure
and magnetic properties of these compounds, with
different types of magnetic \cite{haule,mazin1} or charge \cite{Tesanovic} fluctuations conjectured
to be at the origin of the pairing mechanism.

The rare-earth iron oxyarsenides belong to a wider class of rare-earth oxypnictides
$Re{\rm O}TPn$ (where $Re$ and $T$ are the rare-earth and transition
metals, respectively, $Pn=\rm{P},\rm{As}$ is the pnictogen
ion, with nominal charge $-3$) which have the same
tetragonal structure of $\rm{ZrCuSiAs}$-type \cite{john74}. In spite of obvious similarities in the overall
electronic structure, those compounds differ in their magnetic properties and their tendency towards
superconductivity.
It is particularly interesting in this respect to contrast the properties of LaOFeAs to
those of its phosphate homologue LaFeOP. Relating the observed differences to relevant differences
in their electronic structure may provide important clues into the nature and the origin of
magnetism and superconductivity in these materials.
The undoped stoechiometric compounds have different magnetic properties:
\LP is non-magnetic \cite{carlo_muSR}, while \LAs undergoes a magnetic ordering transition
at a temperature of 134~K \cite{Kamihara-LOFA} preceded by a structural transition at
a slightly higher temperature \cite{cruz}.
Their transport properties above the magnetic transition temperatures are also different,
with the electrical conductivity of \LP being substantially larger than that
of \LAs \cite{Kamihara-LOFA,kamihara-LOFP2}.
%
These compounds also differ significantly in their superconducting properties.
While undoped \LAs is magnetic and non-superconducting,
the situation in the absence of doping for the parent compound \LP is currently
somewhat controversial. A superconducting transition in the range 4-7~K
was originally reported\cite{Kamihara-LOFP}, and recently confirmed in
Ref. \onlinecite{kamihara-LOFP2}, but questioned in Ref. \onlinecite{0805.2149}
which found no superconductivity 
above 0.35 K.
At any rate, electron doping suppresses the antiferromagnetism in \LAs and leads
to a superconducting phase with $T_c$ as high as 26~K, while it
does not induce such drastic changes in the phosphate compound, in which it causes
only a moderate increase of $T_c$ up to 8~K \cite{kamihara-LOFP2}.

The electronic structure of the \LP compound, in particular its Fermi surface (FS),
has been investigated theoretically by Leb\`egue \cite{lebegue}.
His results can be outlined as follows: the \LP FS consists of two ellipsoidal electronic
cylinders centered at the $M-A$ line at the Brillouin zone (BZ) edge, and of two hole cylinders
centered at the BZ center $\Gamma$-Z line, together with a single hole pocket with three-dimensional
(3D)-like dispersion at the $Z$ point.
A basically identical shape of the FS has been proposed in
several theoretical papers \cite{boeri,singh,mazin2} for \LAs,
with electron doping resulting in the disappearance of the 3D-hole pocket, which gets filled,
thus changing the FS topology and increasing the two-dimensional character upon doping.
This topological change and reduction of effective dimensionality has been suggested to be at
the origin of the different types of magnetic fluctuations observed in the
stoichometric and doped \LAs compounds, respectively\cite{mazin1,mazin2}, and
possibly also responsible for the appearance of superconductivity in the doped compound.
It is important to note, in this respect, that most electronic structure calculations
in Refs.~\cite{boeri,singh,mazin2} have been carried out at values of the lattice constants
and internal parameters (the Fe-As and La-O interplane distances) which were obtained
by relaxing the structure within LDA/GGA.
As pointed out very recently by Mazin {\it et al.}\cite{mazin2}, rather small changes in
the \LAs structural parameters, particularly, in the Fe-As interatomic distance, may lead to
the disappearance of the 3D hole pocket from the FS of stoichometric LaOFeAs.
Furthermore, internal
parameters of the \LAs experimental crystal structure are rather poorly
reproduced by theoretical LDA/GGA calculations, with the Fe-As interplane distance being
underestimated by about 6\% \cite{pickett}.

For these reasons, a comparative study of the \LP and \LAs compounds is especially relevant
in order to clarify the relations between electronic properties and the electronic structure
of these materials.
We {\it do know} that stoichometric \LP and \LAs differ substantially, particularly in properties
directly related to the low-energy electronic structure such as magnetic ordering, transport or superconductivity.
Therefore, one may hope and expect
that electronic structure calculations carried out at judiciously
chosen structural parameters should be able to capture this difference.
In this article, we carry out a detailed first-principles investigation of
\LAs and \LP aimed at identifying relevant differences in their electronic structures while relating them
to the small but significant differences which exist between the \LAs and \LP lattice structures.
We show that electronic structure calculations at the {\it experimental} lattice parameters and
internal positions result in the Fermi surfaces of \LP and \LAs having different shapes,
with the 3D hole pocket present in the case of \LP, but not in the case of LaOFeAs. These changes in the FS
topology are due to shortening of the iron-pnictogen bond-length in \LP as compared to LaOFeAs.
We identify the sensitivity of the electronic structure to this bond-length as being due
to the rather high degree of covalency associated with the iron-pnictogen bond, and
illustrate this point by constructing appropriate Wannier functions.

The paper is organized as follows. We briefly outline our calculational approach in Sec. \ref{sec:comp-det}.
The overall electronic structure of \LAs and \LP is discussed in Sec. \ref{sec:BS-large}. Low-energy
aspects of this electronic structure and Fermi surfaces are described in Sec. \ref{sec:low-BS}, and
the differences between the two materials are illustrated by a study of the sensitivity
of these low-energy aspects to the Fe-pnictogen distance. Then, in Sec. \ref{sec:wannier}, we perform
a Wannier function construction which illustrates this sensitivity and relates it to the
covalency of the iron-pnictogen bond. Finally, in Sec. \ref{sec:width} we study the influence of the 
structural properties on the width of the Fe- bands and its material dependence.

\section{Computational details\label{sec:comp-det}}

The first principle calculations of the band-structure and Fermi surface of these materials
presented in Sec. II and Sec. III were performed using the
Full-Potential APW+local orbitals method as implemented in the Wien2k\cite{wien2k} code. We considered 800
$\mathbf{k}$-points in the BZ and checked that all the properties presented in this article were converged
with this mesh. The results that we present here have been obtained within the
Local Density Approximation (LDA)\cite{LDA}  to the exchange-correlation potential but it is
worth mentioning that our conclusions do not vary if we use the generalized-gradient approximation (GGA)\cite{GGA}
instead.
We have also performed a calculation using the Quantum-ESPRESSO
package \cite{pwscf}, with the computational details like in
\onlinecite{kuroki}, and constructed maximally localized Wannier functions
\cite{maxloc} by using the code developed by Mostofi et al.
\cite{wannier.org}. Last but not least, we have performed an
Nth order muffin tin orbital construction for the Fe-d orbitals \cite{nmto}.

\section{Band structure of \LAs and \LP: comparison on a large energy-scale\label{sec:BS-large}}

\begin{figure}[h]
\begin{centering}
\includegraphics[clip,scale=0.82]{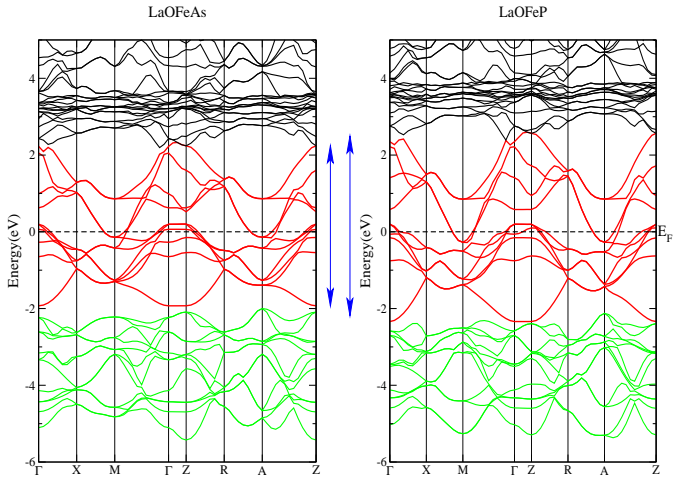}
\par\end{centering}

\caption{Band structure of LaOFeAs (left) and LaOFeP (right). Red and green
bands have mainly Fe-3d and Pnictogen/Oxygen-p character respectively. The colors
are guides to the eyes in order to facilitate comparison of the relative positions and widths of the
bands ( see also blue arrows).\label{fig:whole-range-BS} }
\end{figure}

In this section, we compare the calculated band structure of \LP and
\LAs at their experimental crystal structure, on a large energy scale covering
the whole Fe, $Pn$ (=As or P) and O bands as shown in Fig. \ref{fig:whole-range-BS}.
The crystal structure parameters were taken from experiments and are given in Table \ref{tab:crystal}.
\begin{table}
\caption{Crystal-structure data for \LAs\cite{cruz} and \LP\cite{Kamihara-LOFP}.
Lattice constants \emph{a }and\emph{ c}, internal positions of the
$Pn$ and La (\emph{z}) in units of \emph{c}, and the vertical
distance of the $Pn$ with respect to the Fe plane ($d_{v}$). \label{tab:crystal}
The vertical distance is related to the z-parameter by 
$d_{v} = (z-0.5) c$, and the angle $\theta$ between the As-Fe bond
and the Fe-plane by 
$tan(\theta) = (2z-1) c/a$.
}
  
\begin{tabular}{|c|c|c|}
\hline
 & \LAs & \LP\tabularnewline
\hline
\hline
\emph{a} & 4.0301 \AA & 3.9636 \AA\tabularnewline
\hline
\emph{c} & 8.7368 \AA & 8.5122 \AA\tabularnewline
\hline
\emph{z}(La) & 0.1418 & 0.1487\tabularnewline
\hline
\emph{z}($Pn$) & 0.6507 & 0.6339\tabularnewline
\hline
$d_{v}$ & 1.32 \AA & 1.14 \AA\tabularnewline
\hline
$\theta$ & 33.2$^{\circ}$ & 29.9$^{\circ}$ \tabularnewline
\hline
$d_{Fe-As}$ & 2.41 \AA & 2.33 \AA \tabularnewline
\hline
\end{tabular}
\end{table}

The obtained bandstructures for both materials are depicted on Fig.~\ref{fig:whole-range-BS}.
In this figure, the colors of the bands are simply guides to the eyes in order
to compare more easily their relative positions
and bandwidths. The bands plotted in red and green
have mainly Fe $3d$ and $Pn$-O $p$ character, respectively. The bands shown in
black correspond mainly to La 4$f$ states which are well above the Fe
bands. The blue arrows in between the two plots show the approximate bandwidth of the
Fe $3d$ bands in each compound. One may observe that the bands that are close to the
Fermi level, which correspond to Fe, have larger bandwidth in \LP
as compared to LaOFeAs. This is a plausible explanation of why the former
is observed to be a better metal (in the sense of smaller
resistivity)~\cite{Kamihara-LOFA,Kamihara-LOFP}. This also indicates a
lesser degree of electronic correlations in LaOFeP.

The difference in bandwidth results from a combination of three factors.
First, from the difference in Fe-Fe distance which is significantly smaller in
LaOFeP. Secondly, P and As have different
electronegativity, which leads to a shift towards more negative energies
of the P-bands in LaOFeP. Finally, the $Pn$ has different vertical distances
with respect to the Fe plane, a quantity which is mainly governed
by the internal $z$ parameter, and may influence the width
of the Fe' s bands through the different strength of the Fe-$Pn$ bond.

\section{Low-energy bandstructure and Fermi surface: great sensitivity to
Fe-Pn distance \label{sec:low-BS}}

\begin{figure}[h]
\begin{centering}
\includegraphics[clip,scale=0.85]{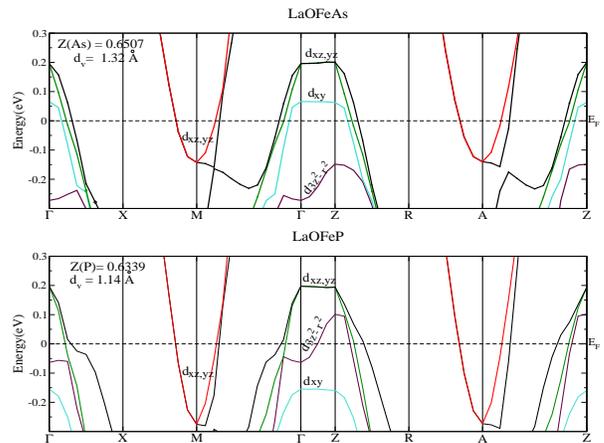}
\par\end{centering}

\caption{Low energy region of LaOFe$Pn$' s band structures with $Pn$= As (up panel)
and P (lower panel).\label{fig:Low-energy-As-P-BS}}

\end{figure}

\begin{figure}[h]
\begin{centering}
\includegraphics[clip]{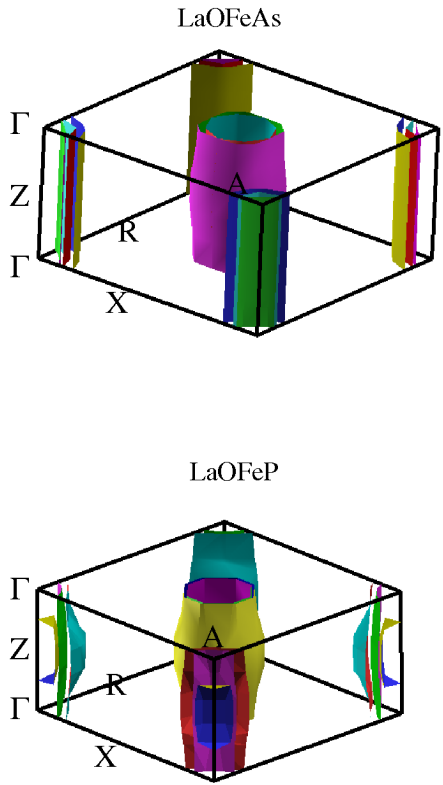}
\par\end{centering}

\caption{Fermi surfaces for LaOFeAs and LaOFeP calculated at their experimental
crystal structure.\label{fig:As-P-FS}}
\end{figure}

In this section, we focus on the very low-energy band structure and
Fermi surface properties of the \LAs and \LP compounds, as
depicted on Fig.~\ref{fig:Low-energy-As-P-BS} and Fig.~\ref{fig:As-P-FS}
(calculated for the experimental crystal structure as given 
in Table \ref{tab:crystal}).

In both compounds, the Fermi level is crossed by five bands
with mainly Fe character. Both materials present two electron pockets
centered at M and two hole pockets centered at $\Gamma$ with mainly
$d_{xz}$ and $d_{xz/yz}$ character. 
(Our choice of the local coordinate system centered at Fe is such that the $x/y$ axes point
towards the nearest-neighbour Fe sites.
With this choice, the $d_{xy}$ orbital is the one that points to
the $Pn$ atom while $d_{x^2-y^2}$ points towards the nearest-neighbor Fe.)
The significant difference between
these compounds comes from the third pocket centered along the $\Gamma\rightarrow Z$
direction. In \LAs this third pocket has mainly $d_{xy}$ character and presents almost no dispersion
along the $z$-direction (it is therefore two-dimensional (2D) in nature). In contrast,
in \LP it consists of a dispersing 3D pocket with mainly $d_{3z^{2}-r^{2}}$
character. 

This difference in the Fermi surfaces may explain the fact that \LAs
is magnetically ordered with better nesting properties than \LP, for which
no magnetic ordering has been reported.
When the situation with the superconductivity of undoped \LP gets eventually
clarified, this difference in the Fermi surfaces may actually prove
to be significant also in this respect.

The explanation for this difference in the low-energy band-structure, as
obtained from first-principles electronic structure methods,
lies mainly in its great sensitivity to the Fe-$Pn$ distance
and to the vertical distance of the $Pn$ to the Fe plane. The reason for this sensitivity
is that, close to the Fermi level, there is a sizeable hybridization between
Fe and $Pn$ states as will be shown later in details.

We checked the effect of doping by using the virtual crystal approximation.
We observe that for the experimental structures there are no qualitative
changes in the low energy band-structure for \LAs, only a shift
of the chemical potential with still three 2D-like pockets crossing the Fermi level.
In contrast, for \LP, the 3D pocket gets
filled and almost disappears from the Fermi surface in the doped compound.
One should note that the bandstructure and FS properties obtained
for the {\it relaxed} \LAs structure (in contrast to the experimental one)
using LDA (also reported by Singh\cite{singh} )
presents features that are very similar to the ones described above for
the \LP material. This is due to the fact that LDA overestimates the Fe-As bonding,
so that the Fe-As distance ($\sim$ 2.4 \AA\  experimentally) in the relaxed structure becomes as small as in the
\LP compound, which is around 2.3 \AA. As a result, the LDA relaxed structure of \LAs
displays a 3D pocket which disappears upon doping.\\

In order to understand better the sensitive dependence of the low-energy
bandstructure on the $Pn$ height with respect to the Fe plane, we have
studied the \LAs system by fixing the lattice parameters
$a$ and $c$ at their experimental values, while varying the vertical position of As all the way from
the experimental value down to a position which is very close to the
one that P has in the \LP material (see Table \ref{tab:crystal}).

\begin{figure}[h]
\begin{centering}
\includegraphics[clip]{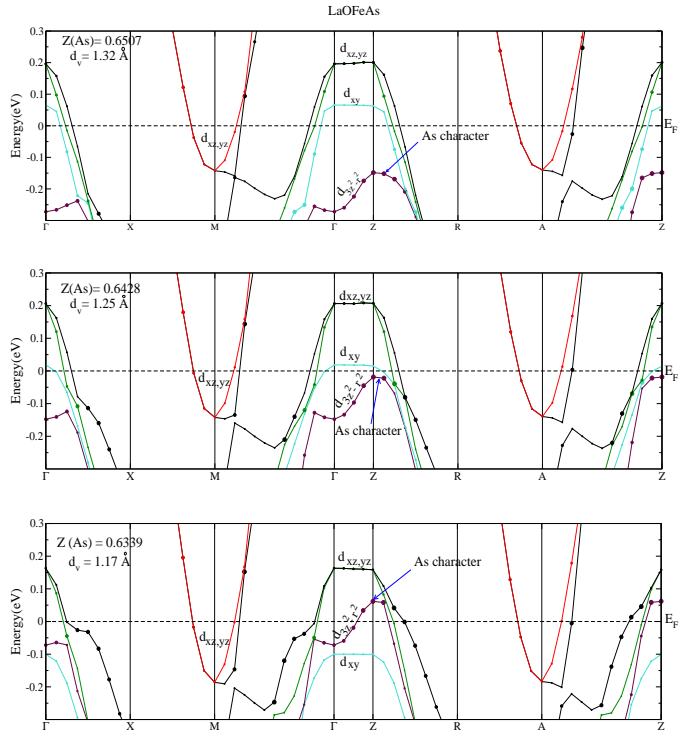}
\par\end{centering}

\caption{Sensitivity of the low energy band structure for LaOFeAs to $d_{v}$
(As' s vertical distance with respect to the Fe plane) for fixed lattice
constants $a$ and $c$. The considered $d_{v}$ values are indicated
in each plot.\label{fig:Sensitivity-BS}}

\end{figure}

The results of this study are depicted on Fig.\ref{fig:Sensitivity-BS}.
We see from this figure that the 2D pocket with $d_{xy}$ character in
the `experimental' \LAs (upper panel) evolves into a 3D pocket
with $d_{3z^{2}-r^{2}}$ character as As is moved closer to the plane (lowest panel).
The circles in the plots indicate  As contribution (mainly As-$p_z$) to the  $d_{3z^{2}-r^{2}}$.  

In a nutshell, one can say that `\LAs evolves into \LP' as the height of the As to
the Fe plane decreases.

\section{Analysis in terms of Wannier functions\label{sec:wannier}}

In the previous section we have noted that the low-energy band structure,
and most notably the topology of the Fermi surface depend very sensitively on
the internal structural parameter $z$(As). Reducing $z$(As) amounts to decreasing the
distance of the As ions to the Fe plane, and changing accordingly
the angle that the Fe-As bond is forming with the Fe-plane.
In this section, we obtain more direct insights into this sensitive
dependence by constructing Wannier functions, which reveal the
strong covalency and hybridization effects associated with the Fe-As bond.
We present calculations done at the experimental structural parameters.

We have used two different types of Wannier constructions.
The first one is a maximally localized Wannier function construction
(MLWF) \cite{maxloc} based on an ultra-soft pseudopotential approach as
implemented in the PWSCF code \cite{pwscf}.
The second one is the N-th Order Muffin Tin Orbitals (NMTO) method
and downfolding technique \cite{nmto}, based on an  LMTO-ASA approach 
\cite{lmto}.
Previous experience has shown \cite{lechermann} that these 
two approaches
give results that are quite comparable to each other.

A crucial issue is the choice of the energy range on which the downfolding is
performed, i.e the set of Bloch bands that the Wannier construction aims
at reproducing. We shall consider two possible choices: a restricted energy
range covering roughly $[-2,+2]$eV around the Fermi level, corresponding to
downfolding on the set of $2*5$ bands with dominant Fe character. We shall call
this the `d-downfolding'. We have also consider downfolding on a much larger energy
range (approximately from $-6$eV to $+2$eV) encompassing all bands with
Fe-d, O-p and As-p character ($2*11$ bands), which we shall denote
by `dpp-downfolding'.

\begin{table}[tbp]
\begin{center}
\begin{tabular}{|r|r|r|}
\hline
Spread &  d-downfolding  & dpp-downfolding \\
(in A) &  & \\ \hline\hline
$z^2$ & 1.78 & 1.02\\
xz & 2.05 & 1.25\\
yz & 2.05 & 1.25\\
xy & 2.36& 1.24\\
$x^2-y^2$ & 1.73& 0.97\\
\hline
As-p & & 1.87\\
As-p & & 1.92\\
As-p & & 1.92\\
\hline
O-p & & 1.26\\
O-p & & 1.28\\
O-p & & 1.28\\
 \hline
\end{tabular}%
\end{center}
\caption{Spread of maximally localized Wannier functions, both for a
model retaining only the Fe-d states and a model that also includes
As- and O-p orbitals. Note in particular the large spread
of the $xy$ orbital, which points towards the As atom.
}
\label{spreads}
\end{table}
In Table II, we display our results for the spatial extension of the
various Wannier functions of LaOFeAs obtained by the MLWF construction,
for both choices of downfolding, as measured by the spread:
$\Omega = \sqrt(<r^2>-<r>^2)$.
The most striking feature is the large values of the spreads, in
particular of the $d_{xy}$ orbital
which points towards the As-atoms, when downfolding is
performed on the restricted set of Fe bands.
Note that the spread of the $d_{xy}$ orbital is actually
comparable to the Fe-As distance.
%
%
The spreads are considerably reduced when performing the
$dpp$-downfolding involving the larger set of bands, as expected.

\begin{figure}[tbh]
  \centering
  \includegraphics[clip=true,width=.28\textwidth,angle=0]
{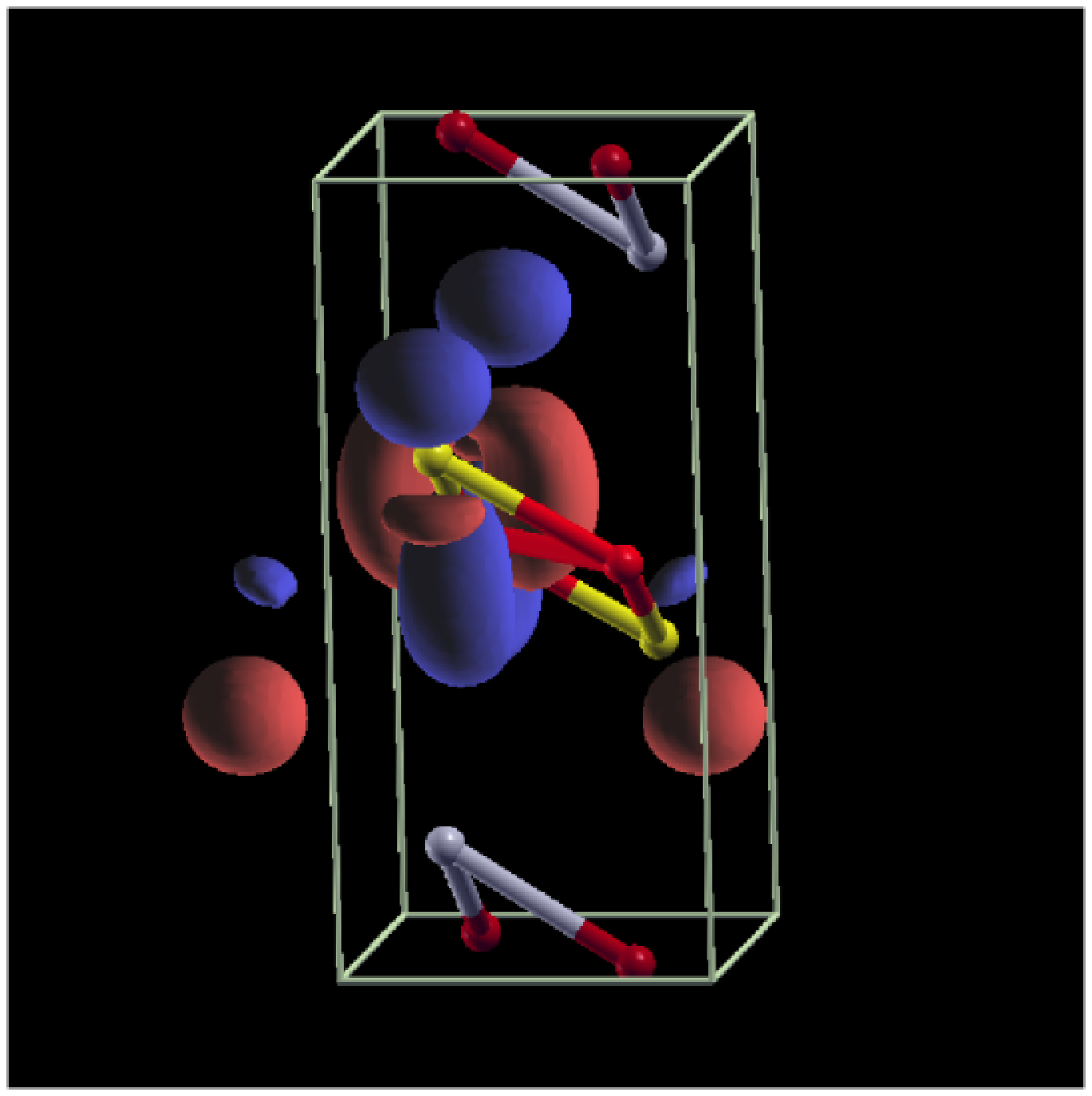}
\\
  \includegraphics[clip=true,width=.28\textwidth,angle=0]
{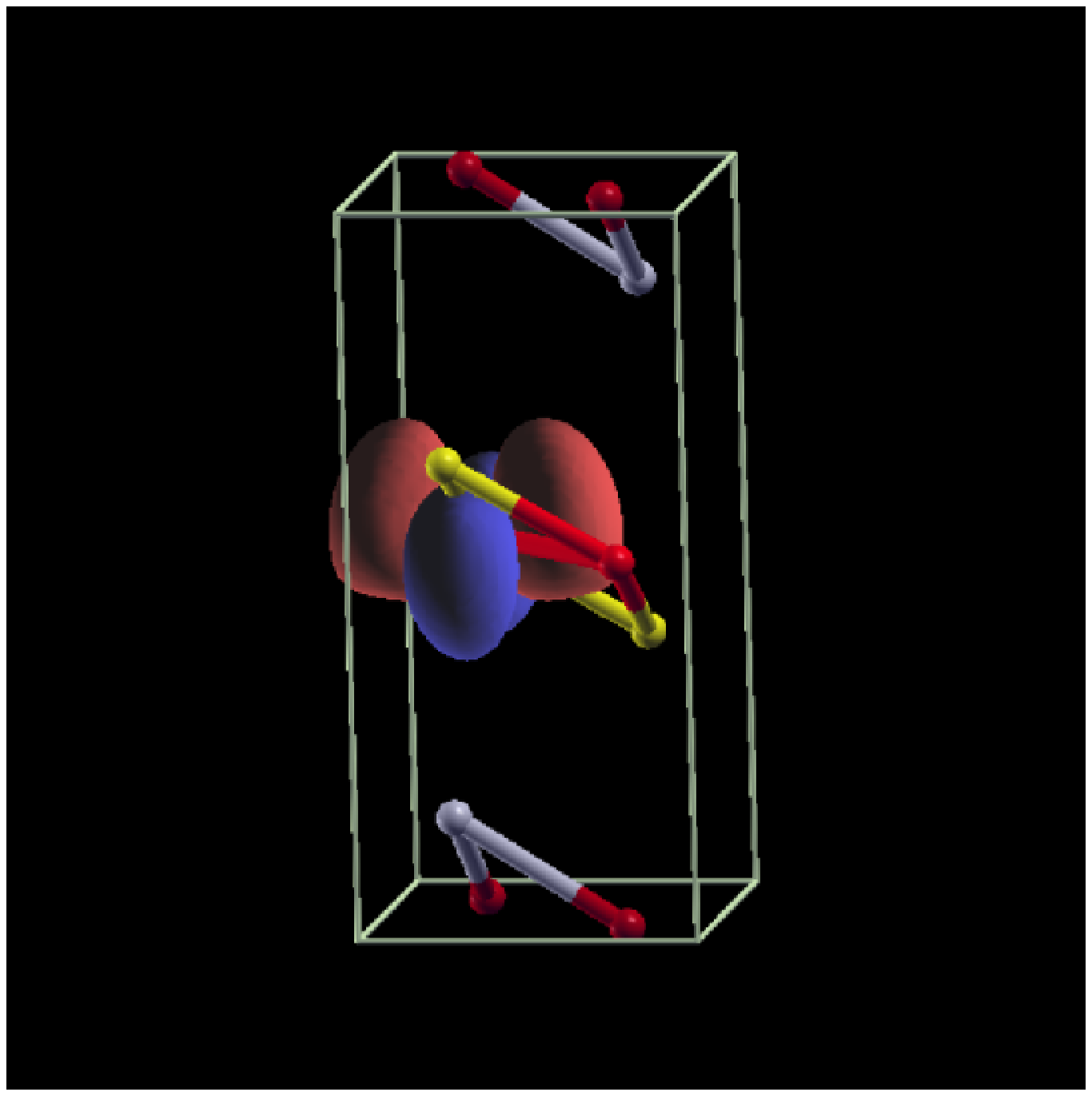}
  \caption{(a) Fe-$xy$ Wannier function for LaFeOAs,
from the maximally localized Wannier construction performed for
the restricted set of Fe-like bands (d-downfolding).
(b) Same orbital character, but within the dpp downfolding, retaining
the full set of Fe-d As-p and O-p bands.}
\label{Wannier}
\end{figure}

In order to visualize this effect, we plot in
Fig.~\ref{Wannier}(a) the isosurface of the
$xy$-like Wannier function. The `leakage' of the Fe state
on the neighboring As atoms is clearly visible.
In Fig. \ref{Wannier}(b) we plot the same isosurface for the same
$d_{xy}$ orbital, but
in the `$dpp$' Wannier construction. As expected, the orbitals are considerably
more localised in this case, and the leakage onto the neighboring
As atoms is suppressed.

In passing, we note that these observations have direct implications
for many-body calculations of these materials performed within the
dynamical mean-field theory (DMFT) framework. Because of the
spatially extended character of the Fe-Wannier functions in the
restricted d-downfolding, it seems that including only
local matrix elements of the Coulomb interaction and local components
of the self-energy would be a rather poor approximation for this
`d-only' model. A full model using dpp-downfolding (hence with more
local Wannier functions) seems like a better starting point for
a DMFT calculation. This observation may help understanding some of the
differences recently reported in DMFT calculations of \LAs
\cite{haule,anisimov}.
However, another important issue is the value of the on-site Coulomb interaction and the efficiency
of screening, which determines  whether these materials should be viewed as in the intermediate or strong correlation regime.

\begin{figure*}[tbh]
  \centering
  \includegraphics[clip=true,width=\textwidth,angle=0]
{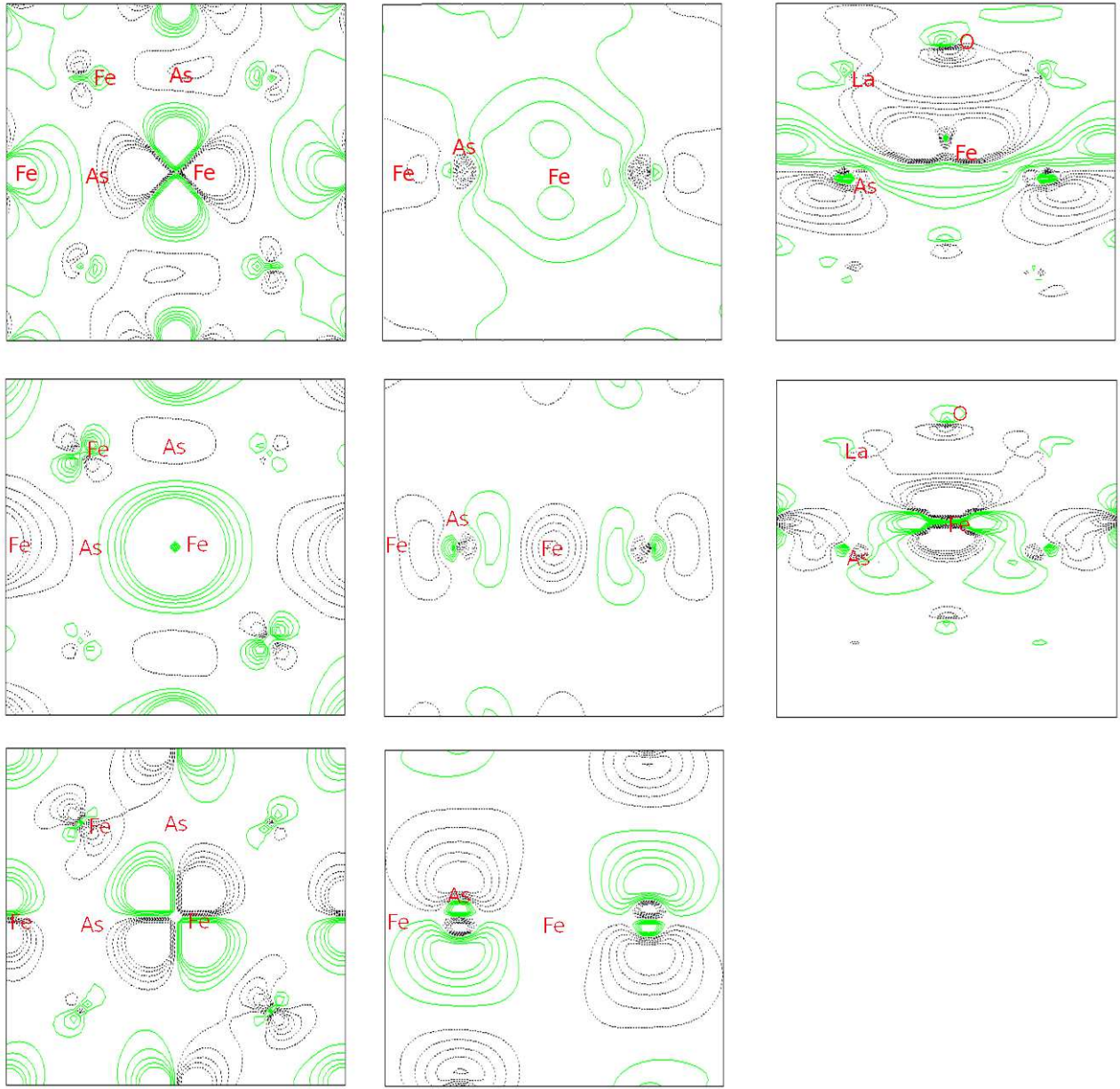}
  \caption{Fe orbitals for the 5-band model for LaFeOAs.
First row: xy orbital.
Second row: $3z^2-r^2$ orbital.
Third row: $x^2-y^2$ orbital.
In the Fe-plane (first column), in the lower As-plane (second
column) and in a plane parallel to the z-axis
containing one La, O, Fe, As per unit
cell (third column). 
In the first two columns the x-axis is along the diagonal
of the plots.
Note that the $x^2-y^2$ orbital
has nodes in the latter plane and is thus not shown in this
representation.
}
  \label{As-d}
\end{figure*}

Finally, in Fig. \ref{As-d}, we display cuts through the isosurfaces of some of
the Wannier functions obtained within the NMTO construction within
d-downfolding.
More specifically, we have chosen to visualise the most
extended orbital (xy), the least extended one ($x^2-y^2$),
and the $3z^2-r^2$ which due to its orientation along the
z-axis plays a special role. Also, as shown in the previous
section, the roles of the xy and the $3z^2-r^2$ orbitals
somewhat interchange concerning the Fermi surface properties
when the $z$ parameter is varied.

We display three specific cuts: the first two
are parallel to the Fe $xy$-plane and correspond to
(i) the Fe-plane, (ii) a plane containing 
the lower layer of pnictide
atoms. The last one is (iii) a cut along a plane parallel to the z-axis and containing
a nearest-neighbour Fe-As bond. It thus cuts through
half of the La, Fe, O and As atoms in the unit cell.
While the first cut mainly gives the orientation of the
Fe orbital, the other two are measures of the leakage of the
iron orbitals onto the As states.
The cuts are represented in the three columns of Fig. \ref{As-d},
the rows show three different Fe-d orbitals: xy, $3z^2-r^2$ and $x^2-y^2$.

The first panel in Fig. \ref{As-d} shows the xy-orbital
 -- the one pointing towards the As atoms -- of the Fe
atom at the center of the plot. Also clearly visible are
the contributions of this orbital on the neighboring four
Fe atoms, as well as parts of the lobes of the four 
next nearest neighbor Fe
atoms.
Note, that the two As atoms do not lie in the plane
represented here, but 1.32 \AA  
 above and below respectively.
Still, they mediate the hopping to the next nearest
Fe atoms, which display more extended xy-character
than the nearest neighbours, consistent with the
observation of strong nearest neighbor hopping.
The second plot in the first row of Fig. \ref{As-d}
shows this same Fe orbital but represented by an isosurface
cut containing the As-plane. The contributions of the
Fe orbital leaking into this plane and in particular onto
the two As atoms (eye-like features) are
clearly visible.
Finally, the cut of this orbital in the plane containing
the z-axis confirms the very extended nature of this
orbital. 

The second row in Fig. \ref{As-d} shows the same cuts, 
but for the
$3z^2-r^2$ orbital, the third row for the $x^2-y^2$ orbital.
Particularly interesting are the second plots in these
rows
which visualise the hybridisation
with the $p_x \pm i p_y$ orbitals of the As atoms,
respectively.
For LaFeOP (not shown) a similar picture emerges.

\begin{figure}[tbh]
  \centering
  \includegraphics[clip=true,width=.28\textwidth,angle=0]
{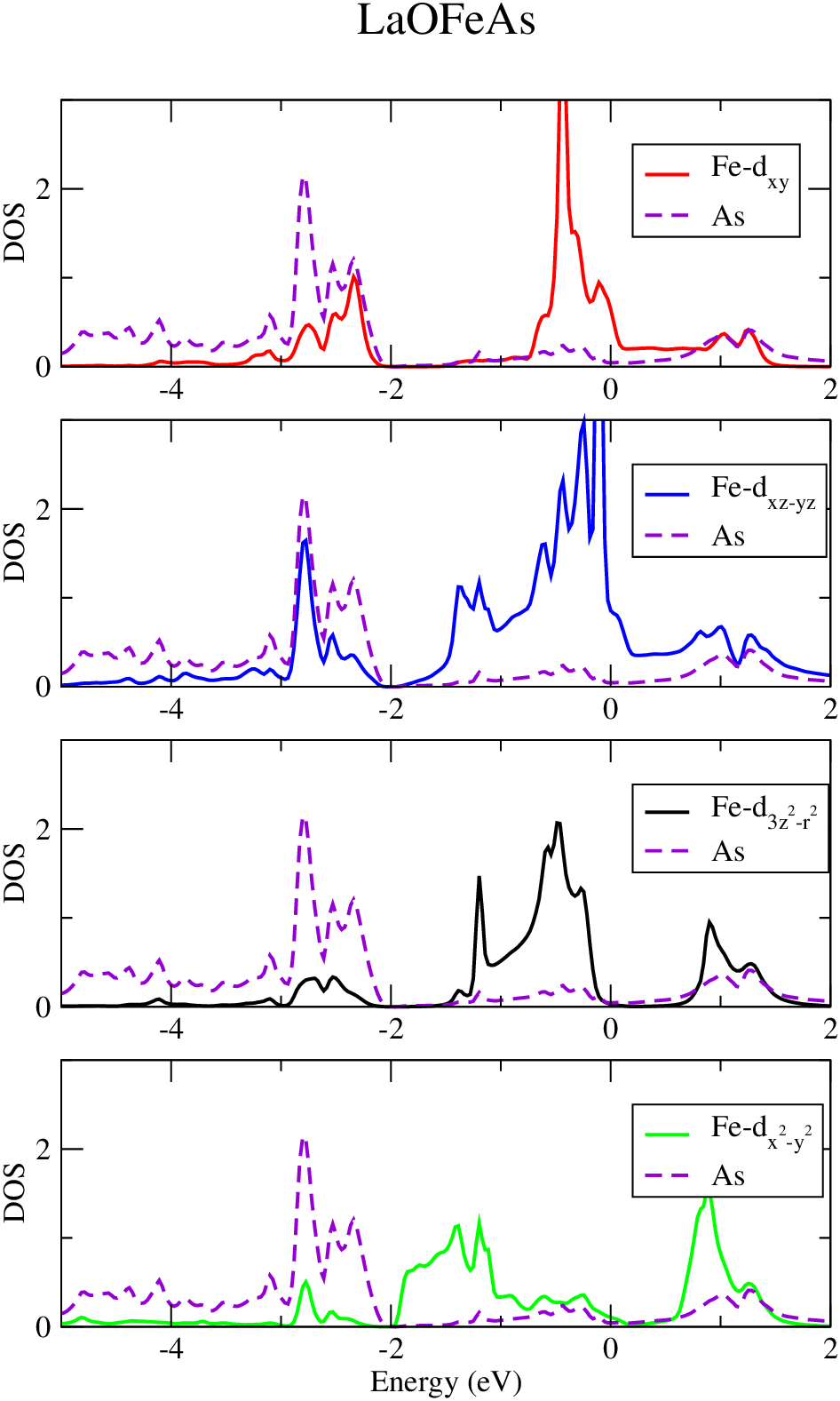}
  \caption{
Partial density of states of the Fe-d orbitals (solid lines), compared
to the As density of states (dashed lines, same curve in all panels).
\label{PartialDOS}}
\end{figure}

The overall comparison of the different orbitals confirms
the result of Table 1 that the $xy$ orbital is by far
the most extended one (followed in fact by the yz and
xz orbitals -- not shown), displaying a tremendous
contribution on the neighboring As sites, in a pancake-like
flat shape.
Fig. \ref{PartialDOS} shows the partial densities of
states (DOS) of the different Fe-d orbitals, in comparison
with the As partial DOS.
The pronounced peaks of the xy and the xz/yz densities
of states in the region of the As-p bands demonstrate
once more that the large spatial extension of these orbitals
is due to their hybridising with the As-p states.

The Wannier analysis can thus be summarized saying
that it shows the large extension of the d-orbitals
when a d-downfolding is used, with the xy- [$3z^2-r^2$]
being the most [least] extended one.
This leakage is consistent with the high sensitivity
of the low energy electronic structure with respect
to the z parameter observed in the previous section.

\section{Width of the iron bands: materials dependence\label{sec:bandwidth}}
\label{sec:width}

Having documented in the previous sections (especially from a comparison
of LaOFeP and \LAs) the importance of covalency and of
the Fe-$Pn$ distance, we finally turn to the influence of structural properties on
the bandwidth of the iron bands, in a broader context.
This question has very recently attracted
attention, with the observation by Zhao {\it et al.i}\onlinecite{zhao} of an apparent systematic
correlation between the superconducting critical temperature of several different As-based
pnictides and the angle formed by the Fe-As bond with respect to the Fe-basal plane.
Specifically, these authors observed that the various materials under consideration have
comparable Fe-As distance (bond-length) $d_{Fe-As}$ while the in-plane lattice parameter
$a$ decreases as $T_c$ increases.
Furthermore, these authors suggested that this may imply a smaller bandwidth $W$ for
materials with highest $T_c$, and hence that superconductivity is enhanced by
increasing correlation strength (increasing $U/W$ ratio). Note, however, that this
would mean a smaller bandwidth with decreasing Fe-Fe distance.

\begin{figure}[H]
  \centering
  \includegraphics[clip=true,width=.28\textwidth,angle=0]
{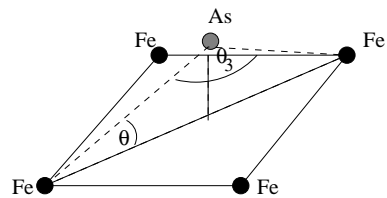}
  \caption{Schematic Fe$_4$As pyramid for LaOFeAs. The $\theta$ and $\theta_3$ angles are indicated.
}
  \label{angles}
\end{figure}

Denoting by $\theta$ the angle formed by the Fe-As bond with respect to the basal plane as sketched in Fig. \ref{angles}
(so that the opening angle at the top of the Fe$_4$As pyramid is
$\theta_3=\pi-2\theta$ in the notations of Ref. \onlinecite{zhao}), the interatomic distances
characterizing the Fe-As unit and the vertical distance of As with respect to the basal
Fe plane are given by:

\begin{eqnarray}
d_{Fe-Fe}=\frac{a}{\sqrt{2}}\,\,,\,\,
d_{Fe-As}=\frac{a}{2\cos\theta}\,\,,\,\, \\ \nonumber
d_v=\frac{a}{2}\tan\theta = c\left(z(As)-\frac{1}{2}\right)
\end{eqnarray}

In order to investigate the dependence of the bandwidth on these structural parameters,
we performed the following studies, which can be viewed as
`numerical experiments':

\begin{itemize}
\item (i) Keeping the lattice constants $a$ and $c$ at their experimental values in
undoped \LAs, we studied how the band-structure evolves as the As is brought
into the plane, starting from its experimental position down to a value comparable to
the height of P in the LaOFeP compound, i.e reducing
$d_v$ or $z$(As) (hence decreasing $\theta$ and increasing $\theta_3$).
The resulting changes in the low-energy
electronic structure corresponding to these calculations have been reported above in
Sec. \ref{sec:low-BS}
\item (ii) Keeping constant the Fe-As distance $d_{Fe-As}$= 2.407 \AA (as well as $c$ at the
same value than above), we reduce the lattice parameter $a$. In so doing, the vertical
distance $d_v$ separating As from the Fe plane increases, resulting in a larger angle
$\theta$ with the basal plane and a smaller angle at the summit $\theta_3$.
\item (iii) Finally, we consider the actual experimental structure of the parent
compounds \LAs, PrOFeAs and SmOFeAs. These three materials have almost identical
Fe-As distance and correspond to decreasing values of $a$, similarly to (ii), but in
contrast to this case they also correspond to decreasing values of the lattice
constant $c$, with the ReO unit getting closer to the FeAs one.
\end{itemize}

\begin{figure}[tbh]
  \centering
  \includegraphics[clip=true,width=.5\textwidth,angle=0]
{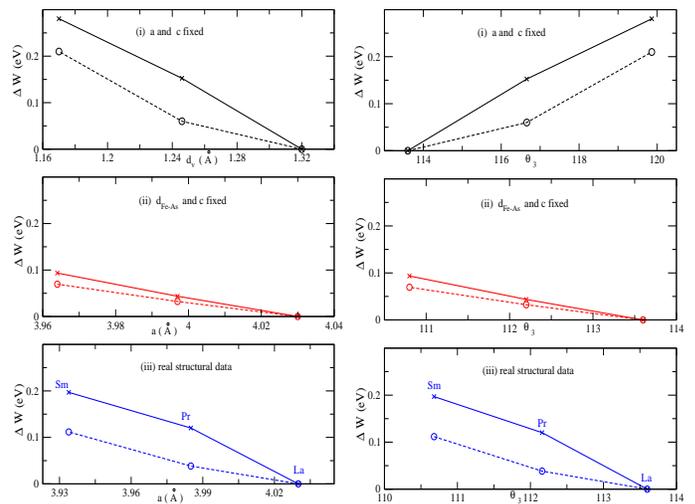}
  \caption{The change $\Delta W$ in the Fe 3$d$ full bandwidth $W$ (solid line) and the bandwidth of its occupied part $W_{occ}$ (dashed line)
  with respect to calculated value for \LAs at its experimental structure\cite{cruz}. 
  These reference values are 4.25 eV for $W$ and 1.95 eV 
for $W_{occ}$. From the top to bottom panels we display the results for the calculations i), 
ii), iii) as explained in the text. Left: $\Delta W$ as a function of $d_v$ or $a$, as indicated. 
Right: $\Delta W$ as a function of $\theta_3$. The structural data for Pr and Sm compounds were 
taken from Ref. \onlinecite{POFA} and \onlinecite{SOFA}, respectively.   
}
  \label{fig:delta-W}
\end{figure}

The results of these investigations are depicted in Fig. \ref{fig:delta-W}, in which we display the
change in both the full bandwidth $W$ of the Fe bands and the bandwidth of the
occupied Fe states $W_{occ}$ (i.e the distance from the bottom of the Fe bands up to
Fermi level). Both quantities provide useful information, but the latter is usually a better
assesment of the relative importance of correlation effects (being related to the typical
kinetic energy).

We note the following trends for the three numerical experiments above:

\begin{itemize}
\item (i) As $d_v$ is reduced for fixed $a$, a systematic increase of the bandwidth
(and of the bandwidth of the occupied states) is observed. This is indeed expected:
the direct Fe-Fe hopping is unchanged, while hybridisation effects with $As$ are increased
as $d_v$ is decreased, hence increasing the hopping through As sites. Thus, in this
case, the bandwidth {\it decreases} as $\theta_3$ decreases (and $\theta$ increases).

\item (ii) In this case where $d_{Fe-As}$ is kept constant, both the occupied and full bandwidth show a smaller effect
 as the $a$ parameter decreases. Hence, there is a slight dependence of the
bandwidth on the angles $\theta,\theta_3$ as compared to i). The reason for this is that the increased
Fe-Fe hopping (due the decrease of $a$) is precisely compensated by the decrease of
indirect hopping through As (due to the increase of $d_v$).

\item (iii) Finally, for the real materials which also have basically a constant
$d_{Fe-As}$, we observe nevertheless a stronger {\it increase} of the bandwidth as $a$
is reduced as compared to ii).
The reason for the difference with (ii) is that the lattice parameter $c$ simultaneously
decreases, and indeed the increase of the bandwidth comes also from
unoccupied bands above Fermi-level, reflecting the influence of the Re-O unit becoming closer.
Hence, for the three materials investigated, the bandwidth  {\it increases} as
$\theta_3$ decreases (and $\theta$ increases), in contrast to (i). The {\it occupied}
bandwidth is even less sensitive to $\theta_3$, again because of the compensating effect observed for (ii).
\end{itemize}

The general conclusion of this investigation is that, as expected physically,
the bandwidth is controlled both by the direct Fe-Fe hopping in the plane and
by the indirect hopping through As (in view of covalency and hybridization effects).
Both the parameters $a$ and $\theta$ (or $\theta_3$) are therefore important.
The bandwidth can in principle behave differently (increase or decrease) as the opening angle
$\theta_3$ is decreased. For the three real materials that we have investigated, the
bandwidth displays a sizable {\it increase} for decreasing $\theta_3$ however, in contradiction
to the proposal made in Ref. \onlinecite{zhao}. This increase is basically related to the
decrease of the c-parameter and the increase Fe-Fe hopping (at constant $c$ the bandwidth experiments a more slight effect).

\section{Conclusions}
 
The iron oxypnictides family of compounds share a similar overall 
electronic structure.
Some important aspects, however, depend very sensitively on small 
structural changes,
particularly on changes of interatomic distances and bond angles 
within the iron-pnictide plane.
Using first-principles full-potential electronic structure 
calculations, we investigate
this sensitive dependence, contrasting in particular \LAs and 
\LP, which display different
transport, magnetic and superconducting properties.
The width of the Fe-bands is significantly larger for \LP, 
indicating a better metal and
weaker electronic correlations. These two materials also 
present significant
differences in their very low-energy bandstructure, when 
calculations are performed at their
experimental crystal structure. Both materials have three 
hole-pockets around the $\Gamma$ point
and two electron pockets around the M-point. However, one 
of the hole pockets changes from
a three-dimensional one to a tube with two-dimensional dispersion 
when going from \LP to LaOFeAs.
These differences are due to the shorter Fe-Fe distance and 
to the shorter distance of the
pnictide to the iron-plane in LaOFeP.
We have shown that the low-energy bandstructure of \LAs evolves towards 
that of \LP as the As atom is lowered from its experimental position 
to a position closer to the
Fe plane.
We argue that the physical origin of this sensitivity to the 
iron-pnictide distance is the
covalency of the iron-pnictide bond, leading to strong hybridization effects.
To illustrate this, we have constructed Wannier functions which are found to
have a large
spatial extension when restricting the energy window to the bands 
with dominant 
iron character.
Finally, we have clarified how the bandwidth changes as one moves along the 
rare-earth series in ReOFeAs.
Since the Fe-As distance remains essentially constant, the Fe-Fe 
distance contracts while the vertical distance
to the Fe plane increases, resulting in a compensating effect and 
rather small changes of the bandwidth. A slight increase of the 
bandwidth of unoccupied states is observed,
which is actually associated with the decreasing distance between 
the Fe-As and Re-O planes.

\acknowledgements 
This work has been supported by the French ANR under project
ETSF, and a computing grant at IDRIS Orsay (project 081393).
We acknowledge useful discussions with J. Bobroff and
K. Nakamura.

%
{}
\end{document}

%% file: def.tex
\def\beq{\begin{equation}}
\def\beqno{\begin{equation}\nonumber}
\def\eeq{\end{equation}}

\def\li6{$^6$\rm{Li}}

